\documentclass[final,3p,twocolumn]{elsarticle}

\usepackage{graphicx}
\usepackage{amsmath}
\usepackage{amssymb}
\usepackage{hyperref}
\usepackage{bm}

\journal{Materials Today}

\begin{document}
\begin{frontmatter}

\title{Harnessing Electrostructural Bistability to Control Defects in Polarizable Materials}

\author[1,2]{Tyler C. Sterling\corref{cor}}
\ead{ty.sterling@colorado.edu, tyler.chase.sterling@gmail.com}
\author[1,3]{Tomke E. Glier}
\author[1]{David Abbe}
\author[1,4]{Rahul Nandkishore}
\author[1,5]{Dmitry Reznik\corref{cor}}
\ead{dmitry.reznik@colorado.edu}
\cortext[cor]{Corresponding Author}

\affiliation[1]{organization={Department of Physics, University of Colorado Boulder}, city={Boulder}, state={Colorado}, postcode={80309}, country={USA}}
\affiliation[2]{
organization={Sterling Independent Study and Research},
city={Killeen},
state={TX},
postcode={76549}, 
country={USA}}
\affiliation[3]{organization={Institute for Nanostructure and Solid State Physics, Universit\"at Hamburg}, state={Hamburg}, postcode={20148}, country={Germany}}
\affiliation[4]{organization={Center for Theory of Quantum Matter, University of Colorado Boulder}, city={Boulder}, state={Colorado}, postcode={80309}, country={USA}}
\affiliation[5]{organization={Center for Experiments on Quantum Materials, University of Colorado Boulder}, city={Boulder}, state={Colorado}, postcode={80309}, country={USA}}

\date{\today}

\begin{abstract}
Structural defects are central to the functionalities of materials. However, conventional approaches to defect control are often slow and imprecise. This work identifies and characterizes a distinct physical phenomenon of electrostructural bistability that provides a rapid and precise pathway for engineering materials with defect-controlled physical properties. A generalized theory of conductivity bistability is developed and experimentally validated to unify two distinct regimes: previously reported electrothermal bistability in weakly polarizable insulators, such as silicon, and electrostructural bistability in strongly polarizable materials, such as TiO$_2$. In agreement with the theory, experiments on TiO$_2$ demonstrate that bulk-homogeneous defect concentrations in electrostructurally bistable materials can be tuned rapidly and deterministically by controlling applied electric current. Abruptly interrupting the current quenches the system into long-lived, room-temperature metastable states with distinct properties. This establishes a practical, rapid, and scalable pathway for defect engineering with direct relevance to nonvolatile memory devices.
\end{abstract}

\begin{highlights}
\item Polarizable insulators exhibit "electrostructural" bistability where carriers come from ionized defects
\item Current heats the sample, creating more defects, increasing the current and so on, leading to runaway
\item Runaway is halted by limiting current -- temperature from Joule heating and applied field cooperate to create a non-equilibrium defect population
\item The enhanced defect population can be quenched in, creating new materials with modified optical and electrical properties
\end{highlights}


\begin{graphicalabstract}
\begin{figure*}
    \centering
    \includegraphics[width=\linewidth]{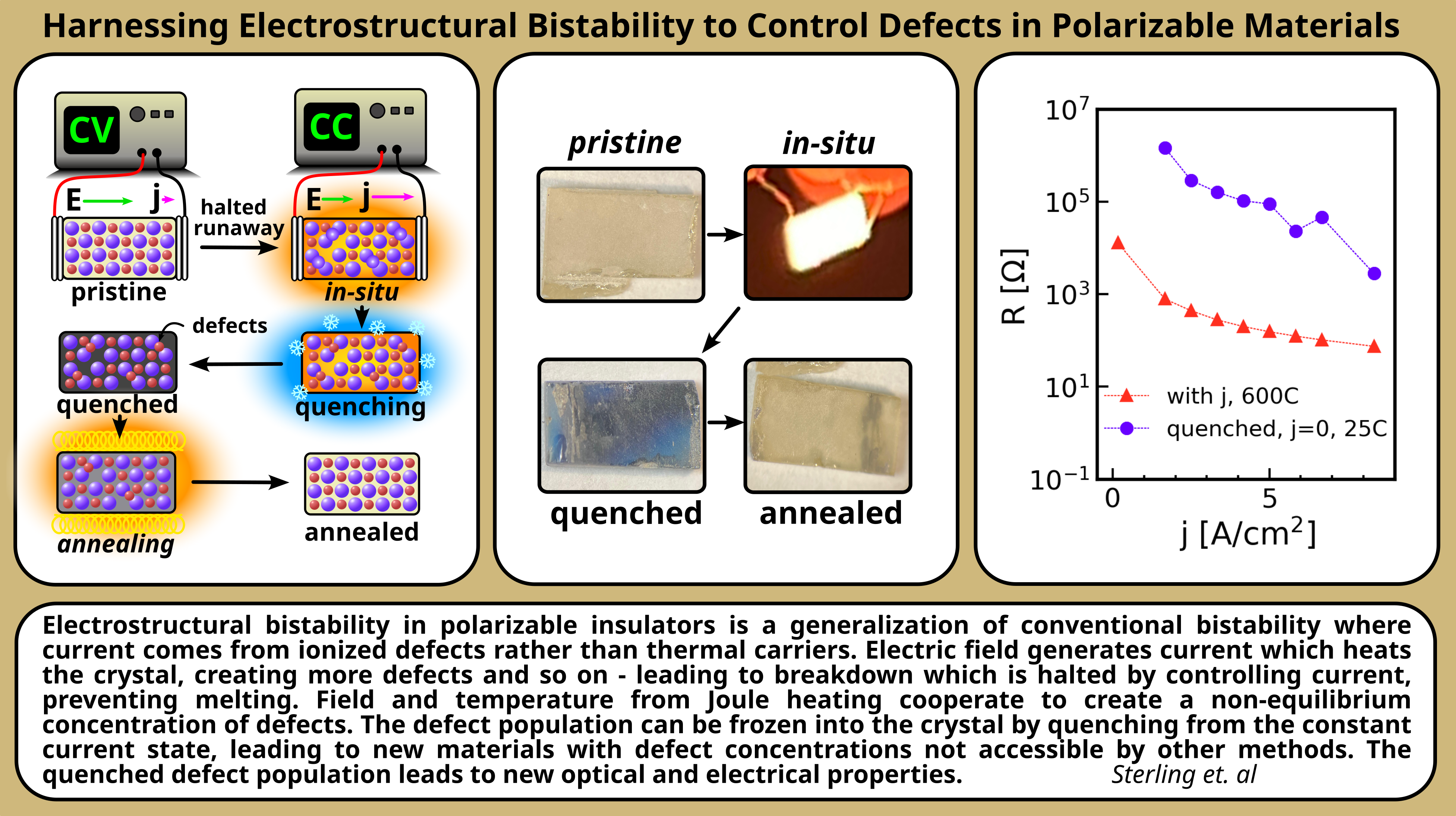}
\end{figure*}
\end{graphicalabstract}

\end{frontmatter}

\section{Introduction}

Learning how to control structural defects is a longstanding central objective of materials science, as defects—vacancies, interstitials, dislocations and extended disorder—govern the fundamental physical properties of solids while also enabling functionalities absent in the ideal crystal. Defects strongly influence electronic structure, lattice dynamics and transport, determine carrier densities, scattering mechanisms, and phase stability, can act as nucleation centers for emergent behavior, and underlie new technologies such as resistive nonvolatile memory (memristors) \cite{alloul2009defects,yang2013memristive,zhu2019anionic,pastor2022electronic,bai2018defect,kalinin2013functional,li2020defects,illarionov2020memristive,szot2011tio2,paz2022universal}. Furthermore, controlled defect populations provide a promising route to materials design: by tuning their concentration and distribution, one can induce new electronic states, enhance ionic and electronic conductivity, and stabilize phases with unique properties. Conventional strategies—typically relying on thermal annealing and chemical reduction/reaction processes—are often slow, difficult to control, and limited in reproducibility and spatial precision \cite{ou2018tuning,li2020defects,xiong2020defect}. Here we demonstrate that exploiting defect-mediated, current-driven \textit{electrostructural} bistability, expected across a broad class of materials, provides a fundamentally distinct and highly effective route to rapid, reproducible, and precisely controlled defect engineering.

\begin{figure*}
    \includegraphics[width=1\linewidth]{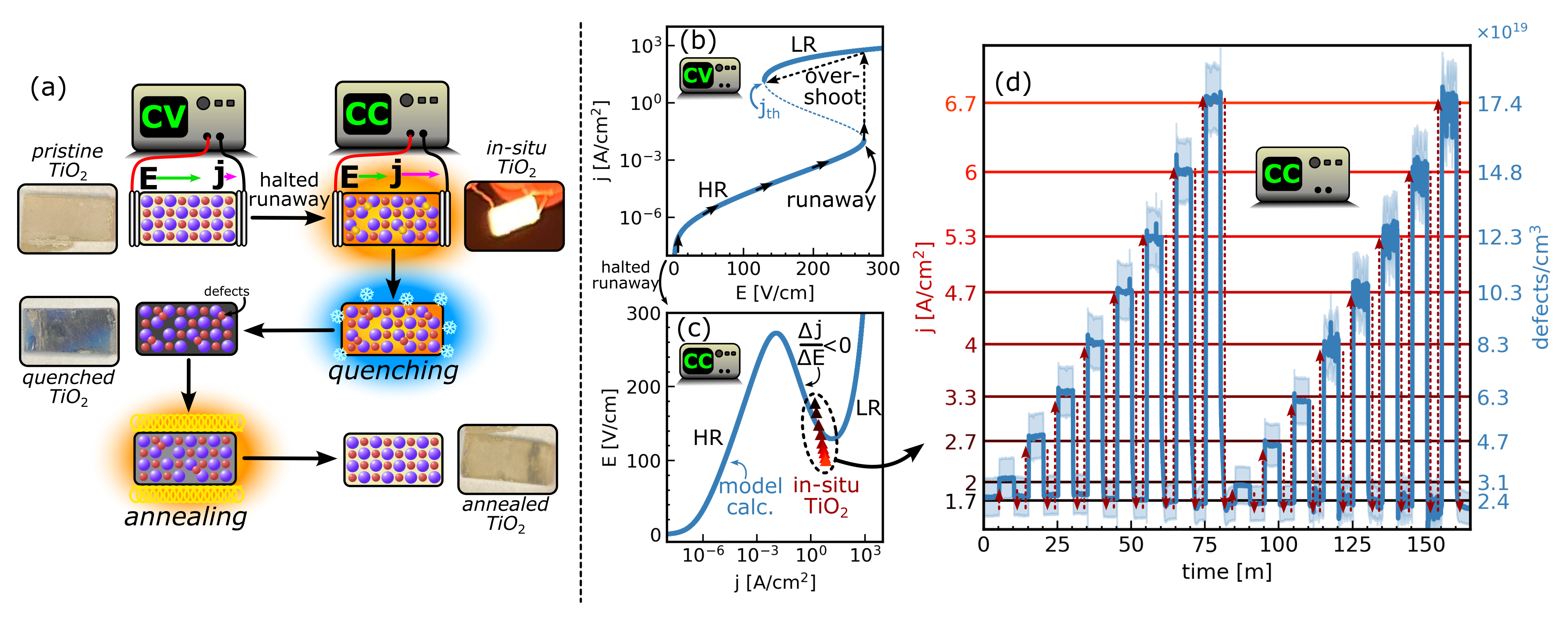}
    \caption{\textbf{Electrostructural bistability driven, current-controlled defect generation in TiO$_2$.} (a) Constant voltage (CV) is applied to a pristine crystal and the field is increased up to the critical value where runaway occurs. Runaway is halted by switching to constant current (CC) mode. Applying current through TiO$_2$ crystals results in defect creation and white visible light production seen as the glow in the photo of the "in-situ" sample. Temperature and field cooperate to enhance the defect concentration which can be quenched in by rapidly cooling the crystal once current is removed. The quenched crystal is opaque but subsequent annealing restores pristine-like clear transparency. (b) Calculated bistable S-shaped $I(V)$ curve: the lower (upper) branch is the HR (LR) state. The dashed line is the unstable state. The calculation assumes model bath temperature $T_b=25^\circ$C and energy lowering parameter $\zeta=6.5\times10^{-4}$ e. (c) Under current control, the entire inverted S-shaped $I(V)$ curve, including the "unstable" NDR region, is single-valued and stable. The triangles are the experimental data from TiO$_2$ in (d). (d) Evolution of defect concentration as a function of time during repeated cycling between low current of 1.7A/cm$^2$ and various higher currents. Horizontal lines are guides to the eye illustrating that the same current yields the same defect concentration, independent of prior history. The shaded regions indicate uncertainty in the defect concentration due to device precision.}
    \label{fig:summary}
\end{figure*}

Both conventional electrothermal bistability (ETB) \cite{altshuler2009jumps} and electrostructural bistability (ESB) introduced here, arise in materials in which electrical resistance drops with increasing temperature. In ETB, resistivity is mediated by conventional thermal excitation of electrons, whereas in ESB resistivity is governed by electric field assisted creation of structural defects that donate mobile charge carriers. In both cases, resistivity takes an Arrhenius form and temperature dependence of electrical power competes with thermal dissipation, producing two stable steady states—a low-temperature/high-resistance (HR) state and a high-temperature/low-resistance (LR) state—separated by an unstable branch exhibiting negative differential resistance (NDR), and giving rise to an S-shaped $I(V)$ curve (Fig. \ref{fig:summary}(b)) \cite{altshuler2009jumps}. 

Here we examine in detail the case of interstitial defects where an atom moves from a crystallographic lattice site to an interstitial site. The occurrence of ESB over ETB is governed by the sensitivity of the activation energy for interstitial defect formation to the applied electric field. This field sensitivity is weak in covalent, small gap materials where the lattice is weakly polarizable and charge carriers are thermally excited band electrons leading to ETB \cite{altshuler2009jumps}. In contrast, in ionic, wide-gap materials, the lattice is strongly polarizable and the electric field dramatically lowers the energy barrier for interstitial defect formation \cite{li2009energetics,abdelouahed2015relevance,moballegh2015electric,schie2018field,kemp2024one,salles2019defect,youssef2017polarizing,el2018effect} resulting in generation of a large concentration of defects at temperatures much lower than expected without the applied field. Much of the current in these materials comes from carriers ionized from defects, leading to ESB: temperature creates defects with concentration enhanced by the applied field, increasing the current which increases Joule heating, further raising temperature, creating more defects and so on until temperature is so high that cooling stabilizes the temperature.

While bistability enables resistive switching in novel devices \cite{ma2002organic,janod2015resistive,ladieu1996depinning,baturina2007localized,ovadia2009electron,illarionov2020memristive,szot2011tio2,paz2022universal}, it is less widely appreciated that ETB and, as we show, ESB also occur in bulk materials. In these systems, however, resistive switching is typically precluded because large Joule heating due to the high current density in the LR state melts the material. This catastrophic failure mode is broadly referred to as thermal runaway. In our experiments, however, automatic switching to current-control mode once the current exceeds a preset limit interrupts the runaway, preventing melting and stabilizing the entire inverted S-shaped $I(V)$ curve, rendering it single-valued (Fig. \ref{fig:summary}(c)). Under both ETB and ESB, systems can be driven continuously through the "unstable" NDR region between the LR and HR states under current-control. 

\begin{figure}
    \centering
    \includegraphics[width=1\linewidth]{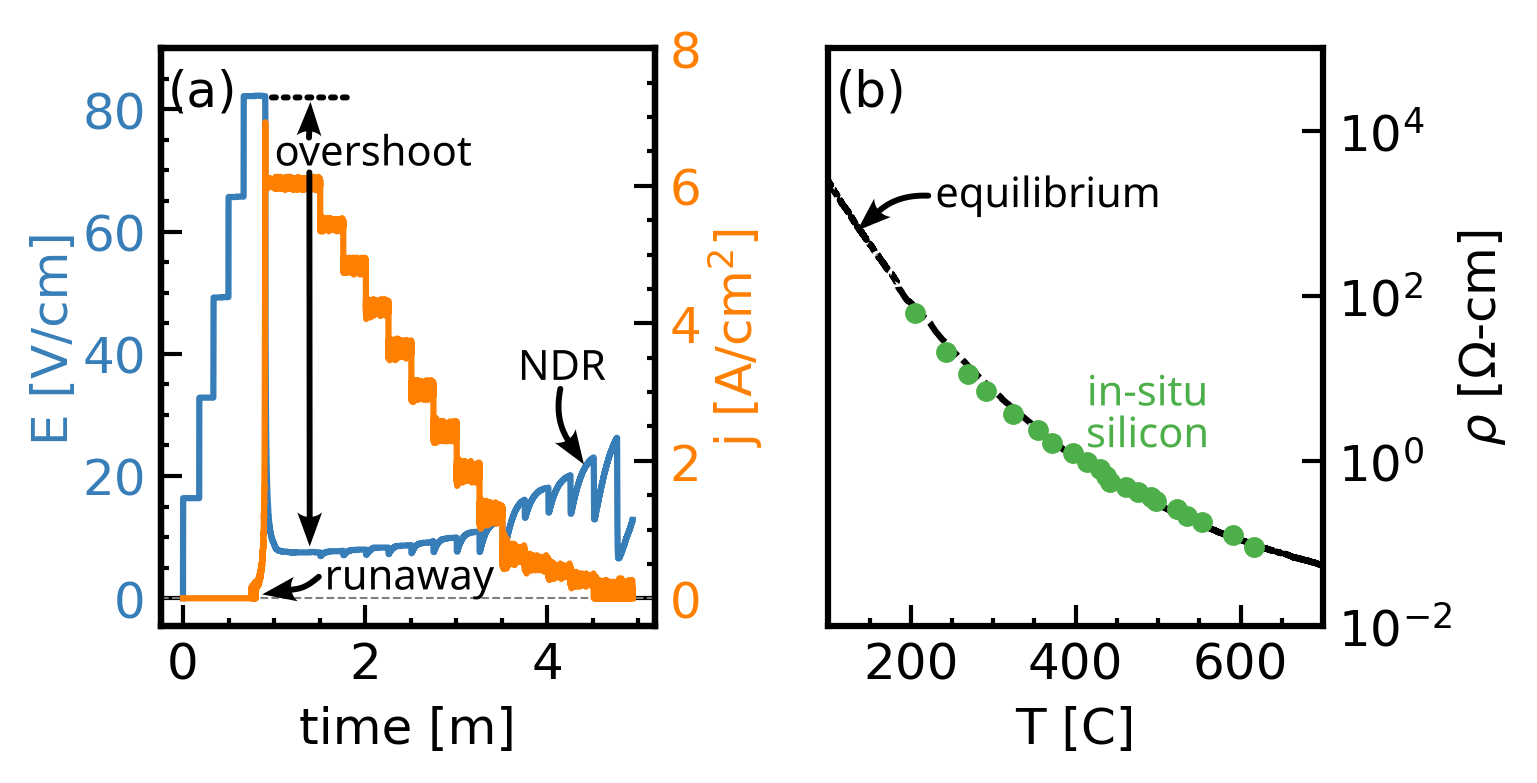}
    \caption{\textbf{Electrothermal bistability in silicon.} (a) Current density and electric field characterizing ETB in silicon with bath temperature $T_b=100^\circ$C. (b) Comparison of the in-situ resistivity of silicon to the resistivity of a crystal heated to the same temperature in equilibrium in a furnace.}
    \label{fig:si_flash}
\end{figure}

We first demonstrate this in bulk silicon, a prototypical material exhibiting ETB. Its nature is the same as established before in films \cite{altshuler2009jumps}. We then demonstrate current-controlled ESB in bulk TiO$_2$, a prototypical material exhibiting ESB. We show here that, remarkably, both types of bistability follow a unified theory, despite their distinct conduction mechanisms. ETB behavior arises directly from the theory of ref. \cite{altshuler2009jumps}, whereas the ESB behavior requires only two key modifications: charge carriers are supplied by ionized defects and the applied electric field reduces the defect formation energy through a Poole–Frenkel–type lowering \cite{frenkel1938pre},  $E(V)=E_0-\zeta V$. Our theory not only exhibits all of the key experimental signatures of both ETB and ESB, i.e runaway, overshoot, and NDR (Figs. \ref{fig:si_flash}(a) and \ref{fig:tio2_flash}(a)), but also includes resistivity-lowering non-equilibrium enhancement of the defect concentration under ESB (Fig. \ref{fig:tio2_flash}(b)), which is a single valued state function of the current density (Figs. \ref{fig:summary}(d) and \ref{fig:tio2_flash}(c)). Defect concentrations produced in the "unstable" NDR region of the S-shaped ESB $I(V)$ curve are only accessible under controlled current. Subsequent quenching from judiciously chosen current densities/bath temperatures freezes in the enhanced defect population, creating new metastable materials with exponentially long lifetimes at room temperature. Specific defect concentrations can be generated deterministically to tune well-controlled optical, electrical, and structural properties that are very different from the unmodified material \cite{sterling2024structural,roy2023structural}. This new ability to deterministically control defect concentration via current has profound implications for many technologies.

\section{Generalized theory of bistability}

There are three key signatures associated with both ETB and ESB as a result of the S-shaped $I(V)$ curve (Figs. \ref{fig:summary}(b-c)): (i) runaway at a critical field, (ii) overshoot before the threshold current is reached, and (iii) negative differential resistance under current control. In addition, ESB develops non-equilibrium defect concentrations (Fig. \ref{fig:tio2_flash}(c)). All of the signatures of the process, including the defect enhancement, can be explained based on the relatively simple concepts of bistability in ref. \cite{altshuler2009jumps}.

Voltage is applied across a material so that a current flows. The current leads to Joule heating which produces energy in the material at a rate $Q_J = V^2 G$, under constant voltage, where $G$ is the conductance and $V$ the voltage. $G$ is different for ETB vs. ESB. The material cools by dissipating energy into a bath (e.g. a furnace) held at temperature $T_b$. The rate at which energy drains from the material due to cooling is $Q_c$. Here, we specialize to the case of a sample in vacuum where heat loss due to cooling is $Q_c = K_0(T^4_b - T^4)$. For simplicity, we assume the coefficient $K_0$ is independent of temperature and carrier/defect concentration. When voltage and bath temperature are held fixed, steady-state is reached and the energy in the material is constant: $\dot U = 0 = Q_J+Q_c$. The sample temperature $\dot T \propto \dot U  = 0$ is also constant.

In covalent semiconductors, the conductance is $G(T) = G_0 e^{-E_a / kT}$ where $E_a = E_g/2$ is the activation energy and $E_g$ the band gap ($E_a \approx 0.5$ eV in silicon). $k$ is Boltzmann's constant. ETB is described by the steady-state solution of the energy balance equation \cite{altshuler2009jumps}:
\begin{equation}
    0 = V^2 G_0 e^{-E_a/kT} + K_0 (T^4_b - T^4) .
    \label{eq:energy_balance_silicon}
\end{equation}
Eq. \ref{eq:energy_balance_silicon} is a special case of the equation studied in detail in ref. \cite{altshuler2009jumps}. They showed that the solution for temperature $T$ indeed exhibits bistability and runaway. They do not consider switching to current control mode.

Here, we generalize this theory to include wide-gap ionic materials like TiO$_2$. For simplicity, we assume the band gap is so large that no electrons can be thermally excited ($E_g\approx 3$ eV for TiO$_2$ \cite{cronemeyer1952electrical,amtout1995optical}) and that all of the carriers come from ionized defects. In line with our simple model, we assume a single defect species. We do not consider the details of specific defect formation in TiO$_2$ which have been well studied before \cite{zhu2014electronic,liborio2008thermodynamics,he2007prediction,li2009energetics}. Rather, we note that under oxygen deficient conditions (e.g. vacuum, inert gas, air), point defects in TiO$_2$ have formation energies $\gtrsim1$ eV \cite{zhu2014electronic,liborio2008thermodynamics} that are further stabilized by high temperature \cite{he2007prediction}, ordering into Wadsley defects \cite{zhang2021andersson,bursill1969new,bursill1971aggregation,sterling2024structural}, and applied field \cite{abdelouahed2015relevance,moballegh2015electric}. 

In general, the conductivity is described by $\sigma = q n \nu$ where $q$ is the carrier charge, $n$ the carrier concentration, and $\nu$ the mobility. In materials like TiO$_2$, $q$ is the defect ionization state and $\delta$ is the number of defects per unit volume. We are interested in the high temperature regime where defects are totally ionized (i.e. $q$ is constant) and the mobility is phonon-limited ($\nu \sim 1/T^\beta$). The defect concentration rate is $\dot n = \Gamma_0 ( n_0e^{-E(V)/kT} - n)$ where $\Gamma_0$ is the defect creation/annihilation attempt frequency, which we take to be temperature dependent for simplicity, and $n_0 = 1.6\times10^{22}$ cm$^{-3}$ is the characteristic defect concentration (see Methods). In the steady-state and with no voltage applied, $\dot n = 0$ and $n = n_0 e^{-E_0/kT}$ which justifies this model for defect rate. The Arrhenius behavior of $n(T)$ in the steady-state dominates over the temperature dependence of the mobility, so we further simplify $G = G(T,V) = G_0 n(T,V)$; indeed this same simplification is used when deducing $G(T)$ for semiconductors. For convenience, we redefine $0 \leq n \leq 1$ as the dimensionless fraction of available defect sites in the whole crystal that are occupied so that $G_0$ has dimensions of conductance. We placed $q$, $\nu$, and $n_0$ into the parameter $G_0$. We still use the term "defect concentration" for $n$. We multiply $n$ by $n_0=1.6\times10^{22}$ cm$^{-3}$ to convert back to absolute units of number of defects per unit volume in TiO$_2$ (see Methods section).

Based on both empirical and theoretical evidence that the applied field enhances defects \cite{schie2018field,kemp2024one,salles2019defect,youssef2017polarizing,el2018effect}, we posit an effective defect formation energy $E(V)=E_0 - \zeta V$ with $E_0$ the bare formation energy and $\zeta \geq 0$ a Poole-Frenkel \cite{frenkel1938pre} like charge parameter that lowers the barrier to defect formation in an externally applied voltage. Such an energy lowering occurs in many oxides: in particular, charged defects, which are most stable in TiO$_2$ \cite{li2009energetics}, form dipoles which interact with external electric fields to further stabilize point defect formation \cite{abdelouahed2015relevance,moballegh2015electric}. In contrast, covalent materials (like silicon) do not couple to the external field since they are non-polar.

The energy balance equation for ESB is
\begin{equation}
    0 = V^2 G_0 e^{-E(V)/kT} + K_0( T^4_b - T^4 ).
    \label{eq:energy_balance_oxide}
\end{equation}
We used the steady-state solution of $\dot n= 0$. See the Methods section for how Eq. \ref{eq:energy_balance_oxide} is solved. When $\zeta=0$, e.g. in a non-polar material, Eq. \ref{eq:energy_balance_oxide} is identical to Eq. \ref{eq:energy_balance_silicon} demonstrating that the nature of bistability is the same in ETB and ESB. The constant voltage $I(V)$ curve in Figure \ref{fig:summary}(b) resulting from solving Eq. \ref{eq:energy_balance_oxide} is S-shaped and qualitatively identical to the solution for ETB \cite{altshuler2009jumps}. In reality, the current under ESB comes from both defect and band carriers, but it is novel to note that a simplified defect-carrier only mechanism of conduction results in an identical form of the theory in ref. \cite{altshuler2009jumps}, yielding bistability and runaway.


In the multivalued region in Figure \ref{fig:summary}(b), there are three branches: the HR branch, the LR branch, and the unstable NDR branch that separates them. Increasing the field from ambient conditions moves the system along the lower branch until the critical field is reached and the system enters the unstable branch, at which point temperature runs away until the system lands on the upper LR branch (Fig. \ref{fig:calc_results}(a)). If voltage were to continue to be held constant, the resulting current density on the LR branch would melt the crystal (Fig. \ref{fig:calc_results}(a-b)). In practice, runaway is halted by switching the power supply to current-control mode once a threshold current is passed. The field abruptly drops from the critical field to the field needed to maintain the threshold current density, resulting in "overshoot" (Fig. \ref{fig:calc_results}(a)).

Once the system is under current-control, the behavior is qualitatively different than under constant voltage. Joule heating under constant current is $Q_J=I^2R(T,I)$. In our model, $R(T,I) = R_0 / n(T,I)$ and the energy balance equation is 
\begin{equation}
    0 = I^2 \frac{R_0}{n(T,I)} + K_0( T^4_b - T^4 ).
    \label{eq:energy_balance_oxide_CC}
\end{equation}
This is valid for both ETB and ESB insulators with $n(T,I)$ different for different carrier sources. If we increase current, we increase temperature, increasing carrier concentration, which \textit{decreases} resistance; there is no feedback enabling "switching". There is only one stable branch of the inverted $I(V)$ curve (Fig. \ref{fig:summary}(c)). Notably, the "unstable" branch becomes stable under constant current. Decreasing current density from the LR moves the system towards HR branch, resulting in an increase in the field: there is negative differential resistance, $\Delta j/\Delta E < 0$ (Fig. \ref{fig:calc_results}(a)). 

\section{Results and Discussion}

\subsection{Electrothermal bistability in silicon}

\begin{figure}
    \centering
    \includegraphics[width=1\linewidth]{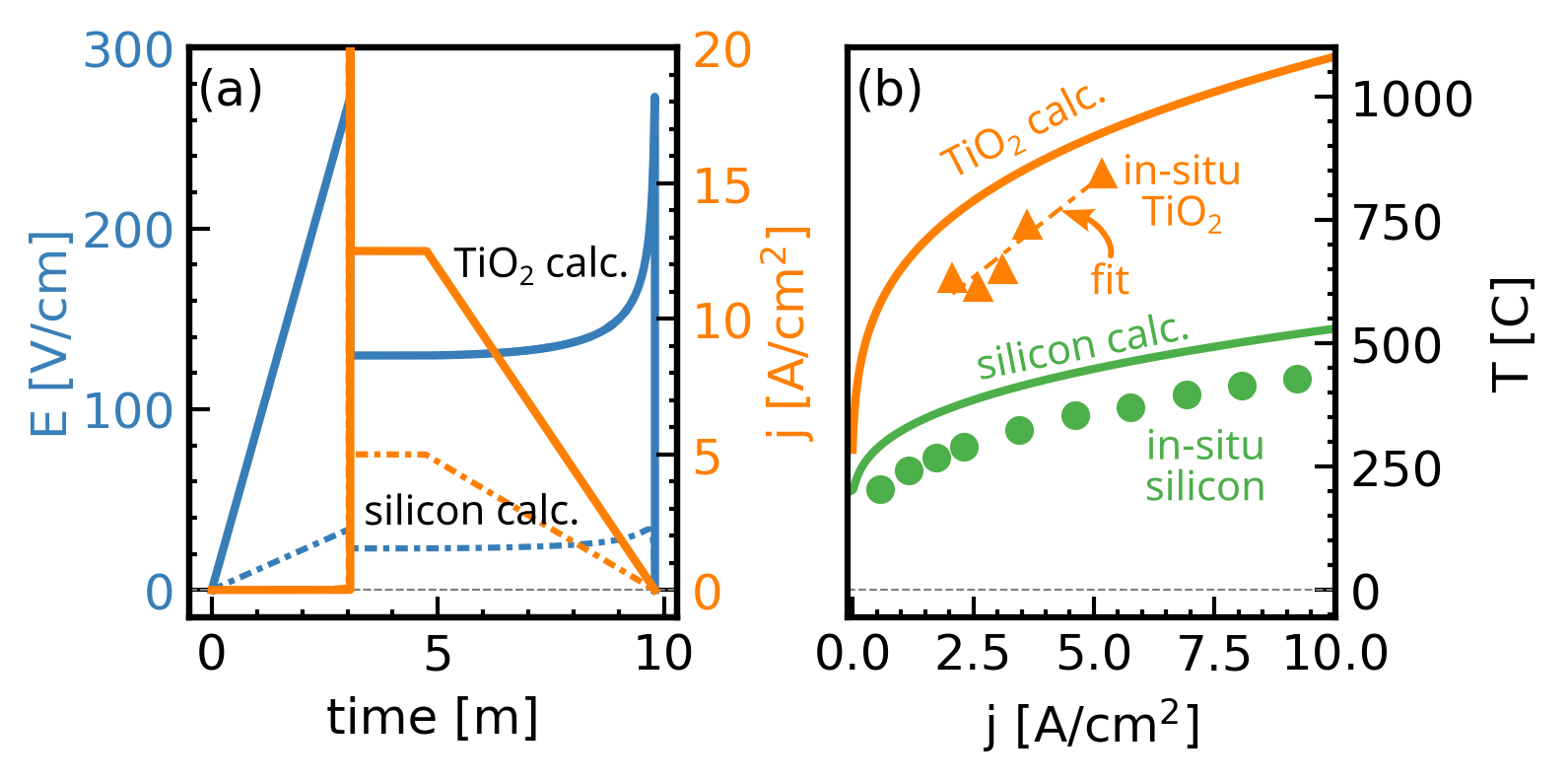}
    \caption{\textbf{Results from calculations.} (a) Simulation of current density and electric field characterizing ESB in TiO$_2$ (solid lines) and ETB in silicon (dashed lines). For TiO$_2$, the model bath temperature and energy lowering parameters are $T_b=25^\circ$C and $\zeta=6.5\times 10^{-4}$ e respectively. For silicon, the model bath temperature is $T_b=-43^\circ$C (see Methods). (b) The in-situ experimental temperature as measured with Stokes/anti-Stokes Raman thermometry (see Methods) compared to calculations. The calculations in (b) used the same parameters as in (a).}
    \label{fig:calc_results}
\end{figure}

We first discuss our measurements of ETB in silicon: a prototypical non-polar, small-gap semiconductor. To initiate and then arrest thermal runaway, we first increased the electric field into the runaway regime (Fig. \ref{fig:si_flash}(a)) and subsequently switched to constant-current control. The current density was then reduced stepwise, allowing the system to reach steady state at each step. This protocol reveals runaway, overshoot, and negative differential resistance (NDR), all consistent with our bistability model (Fig. \ref{fig:calc_results}(a)).

In these and all other measurements presented here, we use electrical resistance as a measure of defect concentration, since defects in insulators are expected to donate extra charge carriers to the conduction band. Thus, if the resistance in the same samples is the same at the same temperature, then the defect concentration is the same. In order to isolate the contributions of carriers donated by defects from those of thermally activated band electrons, it is crucial to accurately measure sample temperatures. Passing current heats the sample, so its temperature is expected to be higher than the bath temperature (Fig. \ref{fig:calc_results}(b)). The in-situ temperature was measured as a function of current density using Stokes/anti-Stokes Raman thermometry (see Methods), which we argue is the most accurate method of measuring temperature in small samples under driven non-equilibrium conditions.

The measured resistivity under applied currents in silicon is indistinguishable from the resistivity at the same temperature when no current passes through the sample (Fig. \ref{fig:si_flash}(b)), demonstrating that transport remains governed by the predicted ETB mechanism, even under high bias, and no defect formation occurs. Because bulk silicon's lattice is non-polarizable, an applied electric field does not couple efficiently to interstitial defects; consequently, the defect population remains independent of applied voltage. Electrical transport therefore arises from only thermally excited band electrons. The electric field required to generate a non-equilibrium population of band carriers in silicon via the Poole–Frenkel mechanism \cite{frenkel1938pre} is substantially larger than the critical field for thermal runaway. For example, impact ionization typically requires $10^4 \lesssim E \lesssim10^6$ V/cm \cite{park1967avalanche,frenkel1938pre}, well beyond the fields at which runaway occurs. Thus, no mechanism exists under our experimental conditions to create a sustained non-equilibrium population of defects or band electrons. 

\subsection{Electrostructural bistability in TiO$_2$}

\begin{figure}
    \includegraphics[width=1\linewidth]{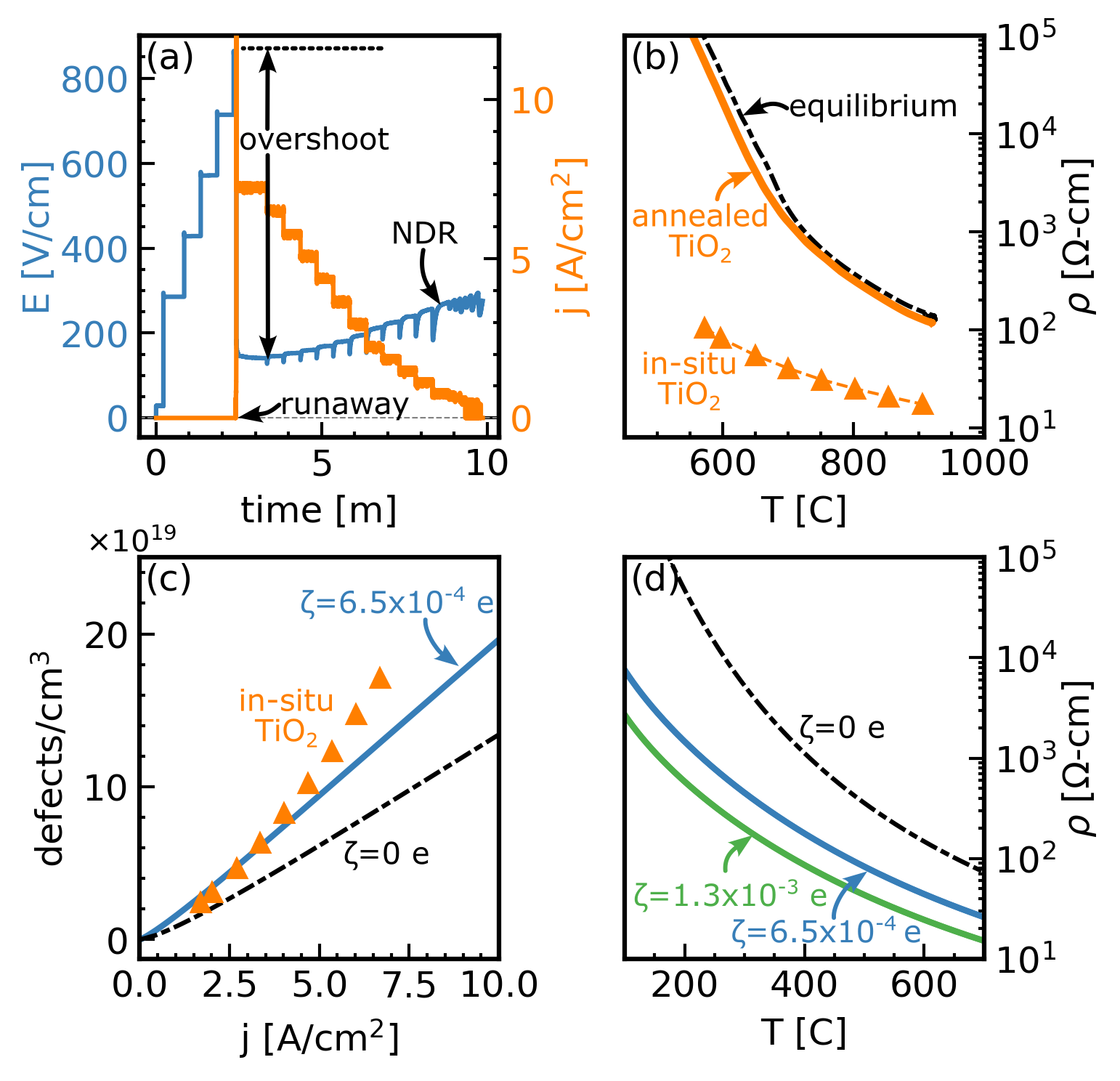}
    \caption{\textbf{Electrostructural bistability in TiO$_2$.} (a) Current density and electric field characterizing ESB in TiO$_2$ with bath temperature $T_b=600^\circ$C. (b) Comparison of the in-situ resistivity of TiO$_2$ to the resistivity of pristine and quenched then annealed crystals heated to the same temperature in equilibrium in a furnace. (c) The defect concentration as a function of current density is estimated from the resistivity as $n(j) = R_0 / \rho(j) \times (1.6\times 10^{22})$ defects/cm$^3$ (see Methods) and compared to calculate defect concentration with $\zeta=0$ e and $\zeta=6.5\times10^{-4}$ e. The calculated resistivity for various $\zeta$ is shown in (d). With $\zeta=0$, only the equilibrium concentration of defects $n_{eq}(T)$ forms.}
    \label{fig:tio2_flash}
\end{figure}

In TiO$_2$, runaway is initiated and halted using the same procedure as in silicon: we increase the electric field into the runaway regime and subsequently switch to constant-current control, revealing runaway, overshoot, and negative differential resistance (Fig. \ref{fig:tio2_flash}(a)). We measured the in-situ temperature (Fig. \ref{fig:calc_results}(b)) and resistivity under current-control using Stokes/anti-Stokes Raman thermometry (see Methods).


 We found that distinguishing characteritic of ESB as opposed to ETB is that the measured resistivity under applied current is orders of magnitude below the resistivity of an untreated TiO$_2$ crystal heated to the same temperature in a furnace without current (Fig. \ref{fig:tio2_flash}(b)). We attribute this to a large, non-equilibrium population of defects Fig. \ref{fig:tio2_flash}(c)) in contrast with silicon where defects do not form. 

For the current densities of our experiments, the in-situ temperature elevated by Joule heating ranges from about 600 to 1000$^\circ$C in TiO$_2$. To isolate thermal excitation from defect creation, we measured resistivity as a function of current density and compared these results to the resistivity of a pristine crystal heated in a furnace while applying only small voltage only for the purpose of measuring resistivity at identical temperature. That is, we compared the resistivity under constant current (where ESB occurs) to that of a field-free pristine crystal (where only thermal effects apply). When heating the crystal to temperature $T$ without applied field, only thermally excited carriers $n_{eq}(T)$ appear. Applied field reduces electrical resistivity by an order of magnitude below the equilibrium value at the same temperatue, which can only be explained by proliferation of structural defects that donate mobile carriers. (Fig. \ref{fig:tio2_flash}(b)).

In further support of the result that ESB generates a non-equilibrium concentration of structural defects, we note that in-situ TiO$_2$ produces intense, white visible light (Fig. \ref{fig:summary}(a)). The measured temperature is too low for the light to be incandescence (Fig. \ref{fig:calc_results}(b)), which should be red, not white. Similarly, the band gap of TiO$_2$ is in the UV range \cite{cronemeyer1952electrical,amtout1995optical}, so the light must not be from band carrier transitions. We deduce that the emission comes from shallow defect state transitions. As discussed below samples quenched from this state are opaque, which indicates a proliferation of mid-gap states that absorb light. Additionally, we note that diffuse neutron scattering points at the presence of substantial oxygen defects in quenched TiO$_2$  \cite{sterling2024structural} and other oxides \cite{roy2023structural}. 

We now show that this behavior is exactly what is expected from the ESB phenomenon described above. TiO$_2$ is a highly polarizable, wide-gap insulator. The most stable defects are charged and form dipoles that couple to the applied field \cite{li2009energetics,abdelouahed2015relevance,moballegh2015electric}. According to our theory, the steady-state defect concentration under ESB is
\begin{equation}
    n(T,V) = n_{eq}(T) \xi(T,V) 
    \label{eq:n_fixed_point}
\end{equation}
where $T$ is the sample temperature, elevated by Joule heating, and $V$ is the voltage needed to maintain the imposed current density. $T$ and $V$ are determined by the solution of the ESB equation under constant current, Equation \ref{eq:energy_balance_oxide_CC}, and $n_{eq}(T) = e^{-E_0 /kT}$ is the ambient, equilibrium defect concentration at temperature $T$, i.e. in the absence of applied field. The field results in an effective lowering of the formation energy \cite{li2009energetics,abdelouahed2015relevance,moballegh2015electric,schie2018field,kemp2024one,salles2019defect,youssef2017polarizing,el2018effect}, increasing the defect concentration by an amount $ \xi(T,V) = e^{\zeta V /kT} $. 

We now discuss the model parameters that could describe the experimental behavior of TiO$_2$. $E_0$ is the “bare” defect activation energy, which is roughly half the defect formation energy. In the absence of electric field (V=0), the defect concentration is the well-known form $n \approx e^{-E_0 /kT}$. Defects in rutile TiO$_2$ have been well studied before and have a formation energy ranging from $\approx 500$ meV to several eV \cite{morgan2010intrinsic,lee2012calculation}. We estimate $E_0 \approx 513$ meV and $\zeta=6.5\times10^{-4}$ e (fundamental charges) for TiO$_2$ (see Methods). Using these numbers in the calculation give reasonable qualitative results. At $V=500$ V, a typical value to initiate the runaway at $T_b\approx600^\circ$C, the energy lowering is $\zeta V \approx 320$ meV, which is reasonable \cite{abdelouahed2015relevance}. Here $\zeta$ is a Poole-Frenkel-like parameter that scales with the defect formation energy. A conceptual picture is as follows: defect formation both creates dipoles directly (e.g. positive cation leaves behind a negative vacancy) and changes the polarization in the crystal by distorting the polar lattice and displacing electronic charge. Under applied field, the the defect formation energy is lowered due to the reduction of the electrostatic and strain energy required to form a defect \cite{abdelouahed2015relevance, youssef2017polarizing}. At $j=5$ A/cm$^2$, the sample is heated to $T\approx750^\circ$C (Fig. \ref{fig:calc_results}(b)) and the field needed to maintain the current density falls to $E \approx 150$ V/cm (Fig. \ref{fig:tio2_flash}(a)). Assume a $1$ cm sample. At this temperature, $n_{eq}(T) \approx 0.003$, i.e. $0.3\%$ of all defect sites are occupied in equilibrium, a reasonable estimate. The field enhancement results in $\xi(T,V) \approx 3.03$, i.e. $303\%$ increase in the number of defects. Then $\approx1\%$ of possible defect sites are occupied with an effective temperature $T\approx1000^\circ$C. 

The defect dynamics are sufficiently slow to allow us to quench the enhanced defect population in TiO$_2$ by removing the current and rapidly cooling the crystal. In our experiments, we quench the crystal by quickly removing it from the furnace immediately after the current is switched off, allowing it to cool rapidly to room temperature. More precise quenching can be achieved using forced air or liquid nitrogen \cite{sterling2024structural}. Quenching a crystal from the current-controlled ESB state produces pronounced changes in both its optical and electrical properties: the quenched crystals become opaque ( Fig. \ref{fig:summary}(a)) and exhibit substantially reduced resistivity \cite{sterling2024structural}, which persists even at room temperature and enables potential practical applications of these treated materials in devices.
The defects generated during ESB appear to be distributed throughout the bulk of the TiO$_2$ crystal. We do not typically observe macroscopic filaments or other spatial inhomogeneities under a 5$\times$ optical microscope. This behavior is consistent with the fact that the low-resistivity state is reached at temperatures above $500^\circ$C (Fig. \ref{fig:tio2_flash}(b)), where defects are sufficiently mobile to diffuse throughout the crystal (see Supplementary Material). It is also consistent with the observed history independence of the $V(I)$ characteristics (Fig. \ref{fig:summary}d). Nevertheless, we cannot exclude clustering or ordering of point defects on the nanoscale. Indeed, previous work has provided evidence for the formation of Wadsley defects in TiO$_2$ \cite{sterling2024structural}, as well as superlattice ordering of oxygen defects in Pr$_2$CuO$_4$ \cite{roy2023structural}.

The defect population generated by ESB can be controllably removed by annealing (see Supplementary Material). To demonstrate this, we placed a quenched crystal in a furnace without an applied electric field and monitored its resistivity as the temperature was gradually increased. Above approximately $500^\circ$C, the resistivity rises rapidly, indicating the progressive removal of defects. After several hours at elevated temperature, subsequent thermal cycling yields a resistivity nearly identical to that of pristine TiO$_2$, indicating that the defects have been effectively eliminated (cf. the annealed curve in Fig. \ref{fig:tio2_flash}(b)). Consistent with this interpretation, the annealed crystal becomes optically transparent, like pristine TiO$_2$ (Fig. \ref{fig:summary}(a)).

Because ESB unambiguously generates defects, one might argue that the resistivity drop simply reflects thermal excitation within a quenched, defect-rich crystal—for example, through changes in the band gap or carrier density. However, when a quenched crystal is heated above approximately $600^\circ$C in a furnace, its defects rapidly anneal out, and its resistivity becomes indistinguishable from that of a pristine crystal (see Supplementary Information). Thus, purely thermal treatment removes rather than creates the defects responsible for the low-resistivity state, making a field-free, high-temperature comparison between quenched and pristine crystals physically uninformative. Taken together, these observations demonstrate that ESB not only generates defects but also drives the crystal into a metastable defect state that cannot be accessed through thermal treatment alone.

\subsection{Relation to flash sintering}


Applying an increasing voltage at constant temperature until the onset of exponential growth in electrical current often induces the enigmatic phenomenon known as flash \cite{raj2021flash,raj2016analysis,raj2012joule,cologna2010flash}.
As in our experiments, once the current limit of the power supply is reached, the system switches to constant-current mode, where the voltage is automatically regulated to stabilize the current at a preset value. During flash, the onset of visible photoluminescence often accompanies a dramatic drop in electrical resistance and powder consolidation into dense ceramics occurs remarkably rapidly and at significantly lower temperatures than in conventional sintering \cite{cologna2010flash}. Flash has also been used to synthesize novel materials such as high-entropy oxides \cite{rost2015entropy,tyagi2024flash}, to create conductors from insulators \cite{sterling2024structural}, and to radically modify structural, electrical, optical, and other properties of materials \cite{sterling2024structural,roy2023structural}. Notably, in single crystals flash preserves crystallinity, as evidenced by sharp Bragg diffraction peaks \cite{sterling2024structural,roy2023structural}. Despite intense research efforts \cite{jongmanns2018generation,jongmanns2020element,raj2012joule,molina2024athermal}, the lack of a clear atomistic mechanism remains the primary obstacle to further progress.

Here we report detailed experiments and calculations on silicon—where flash had not previously been explored—and on TiO$_2$, a prototypical flash material. The experiments demonstrated significant nonequilibrium Joule heating that occurs during these experiments that we reliably measured using Raman scattering. Our results demonstrate a bistability-induced exponential increase in power dissipation in both silicon and TiO$_2$ that could be interpreted as evidence of flash in both systems if Joule heating is underestimated. However, we show that this behavior arises from fundamentally different mechanisms: purely thermal ETB in silicon and ESB dominated by defect formation in TiO$_2$. Because only the latter reproduces the phenomenology associated with flash, our results imply that  materials exhibiting ETB cannot undergo flash sintering and related flash phenomena and those exhibiting ESB can. Our results establish flash in ionic, wide-gap insulators as a current-controlled LR state where interstitial defects are constantly generated and recombine while their mobility is greatly enhanced. 

We propose that the most straightforward way to distinguish between these two classes of insulators is to induce the bistability transition by applying an increasing voltage at a fixed environmental temperature while monitoring the actual sample temperature via Raman scattering. The resulting high-current resistivity is then compared to the low-current resistivity measured at the same sample temperature (achieved by externally heating the furnace to match the Joule-heated Raman-measured temperature). If the resistivity values are identical in both cases, as observed in silicon, defect formation is absent, making flash sintering impossible. Conversely, if the high-current resistivity is lower than the low-current value at identical, accurately measured sample temperatures, the material is a viable candidate for flash sintering.

It is worth noting that electric-field-induced defect formation without exponential runaway has also been reported in metals \cite{das2026loss,bamidele2024current,bamidele2024flash}. Bistability requires positive feedback via the conductivity, which is expected for both thermal excitation in covalent semiconductors and defect generation in ionic insulators but not in metals where conductivity \textit{decreases} with increasing temperature. Establishing possible connections between these effects and the mechanism discussed here is an active area of research and lies beyond the scope of the present work.

\section{Conclusion}

It is widely believed that inducing current runaway in insulating materials produces irreversible structural changes typically interpreted as damage. However, earlier studies of flash phenomena \cite{cologna2010flash} demonstrated that runaway can be arrested before such irreversible changes occur, allowing the material to be annealed back to its original structure once the current is removed. These studies also showed that rapid quenching from the LR state associated with arrested runaway can stabilize interstitial defects in a wide range of materials \cite{sterling2024structural}. Here we showed that the concentration of these defects can be precisely controlled by the combination of electrical current and sample temperature and develop a model of this phenomenon using the framework of ESB. As illustrated in Figure \ref{fig:summary}(d), the current density uniquely determines the reversible creation of interstitial defects. When the current density is reduced, the defect population relaxes to the non-equilibrium steady state value corresponding to that current density, eliminating any excess defects generated at higher current. Importantly, the defect concentration depends only on the instantaneous current density and sample temperature (which is determined by current density and bath temperature) and is independent of the path taken to reach that state. This behavior enables precise and reversible control of the defect population, which can subsequently be stabilized by quenching as described above (see Supplementary Material for details).

We estimate the defect concentration created by ESB from the measured resistivity by comparison with model calculations. In the model for ESB, resistivity is expressed as $n(j)=R_0/\rho(j)$, where $\rho(j)$ is the resistivity and $n$ is the dimensionless defect fraction. The current and resistivity scales of the model are chosen so that the theoretical results are in reasonable agreement with the measurements (see Methods). Notably, the dependence of defect concentration on current density obtained from experiment closely follows the model prediction (Fig. \ref{fig:tio2_flash}(c)). The inferred relation $n(j)=R_0/\rho(j)$ is nearly linear over most of the current range, with slight saturation at low current densities.

These properties enable calibration of defect concentration as a function of bath temperature and current density, thereby enabling reproducible modification of a parent insulating material by introducing controlled concentrations of interstitial defects through electrostructural bistability (ESB).

We propose that a similar mechanism underlies switching in oxide-based memristors, albeit with important distinctions. In particular, switchable memristor materials such as TiO$_2$ typically exhibit localized conducting filament formation rather than a homogeneous LR bulk state \cite{szot2011tio2,illarionov2020memristive}.
We attribute this behavior to lower effective operating temperatures in conventional device architectures, arising from efficient heat dissipation through surrounding structures. Our results further indicate that control of thermal boundary conditions and current density enables stabilization of a bulk LR state in such systems, in place of filamentary conduction.
Detailed exploration of this regime lies beyond the scope of the present work.

In summary, we developed and experimentally validated a generalized theory of electric bistability with the central finding that the defect concentration in TiO$_2$ is a unique and reversible state function of the current density. Once electric current runaway is initiated and arrested, the system enters a steady-state constant-current regime in which a well-defined defect concentration determined by temperature and current density is established. Increasing the current increases the defect population, while reducing the current returns the defect concentration to the original steady-state value (Fig. \ref{fig:summary}(d)). This mechanism, governed by electrostructural bistability, provides a new pathway to create highly doped and potentially switchable and tunable metastable insulating materials—a class of materials that cannot be realized through conventional synthesis routes.

\appendix 

\section{Methods}

\subsection{Stokes/anti-Stokes Raman Thermometry}

Determining the actual sample temperature under constant current conditions is challenging \cite{raj2012joule,charalambous2018investigation}. Lack of thermal equilibrium with the environment renders solid-state temperature probes ineffective for accurate measurement and luminescence and volume emission makes black-body radiation an unreliable indicator of the actual temperature \cite{charalambous2018investigation}. Here, we used Raman scattering to accurately measure temperature of the samples in-situ by applying the principle of detailed balance \cite{yang2017novel,pellatz2021relaxation,kash1985subpicosecond}. Raman spectra were obtained using a 532 nm laser with samples in air. We used a single-stage McPherson Raman spectrometer with an 600 groove/cm grating with a notch filter and a liquid-nitrogen cooled CCD detector. 

We read off temperature from the ratio of intensities on the Stokes (S) and anti-Stokes (AS) sides, $I_S$ and $I_{AS}$. They are related by the fundamental principle of detailed balance:
\begin{equation}
    g\frac{\textrm{I}_{AS}}{\textrm{I}_{S}} = \frac{\left(\omega_0 + \omega_{ph}\right)^4}{\left(\omega_0 - \omega_{ph}\right)^4}\textrm{e}^{-\hbar\omega_{ph}/kT_{ph}},
\label{eq:detailed-balance}
\end{equation}
where $\omega_0$ is the laser frequency, $\hbar\omega_{ph}$ is the phonon energy, $k$ is Boltzmann’s constant, $T_{ph}$ is the phonon temperature, and $g$ is typically equal to 1. Introducing $g$ allows us to correct for the systematic error due to imprecise spectrometer calibration. We fixed $g = 0.9$ to set the phonon temperature equal to $25^\circ$C at room temperature.

\begin{figure}
    \centering
    \includegraphics[width=1\linewidth]{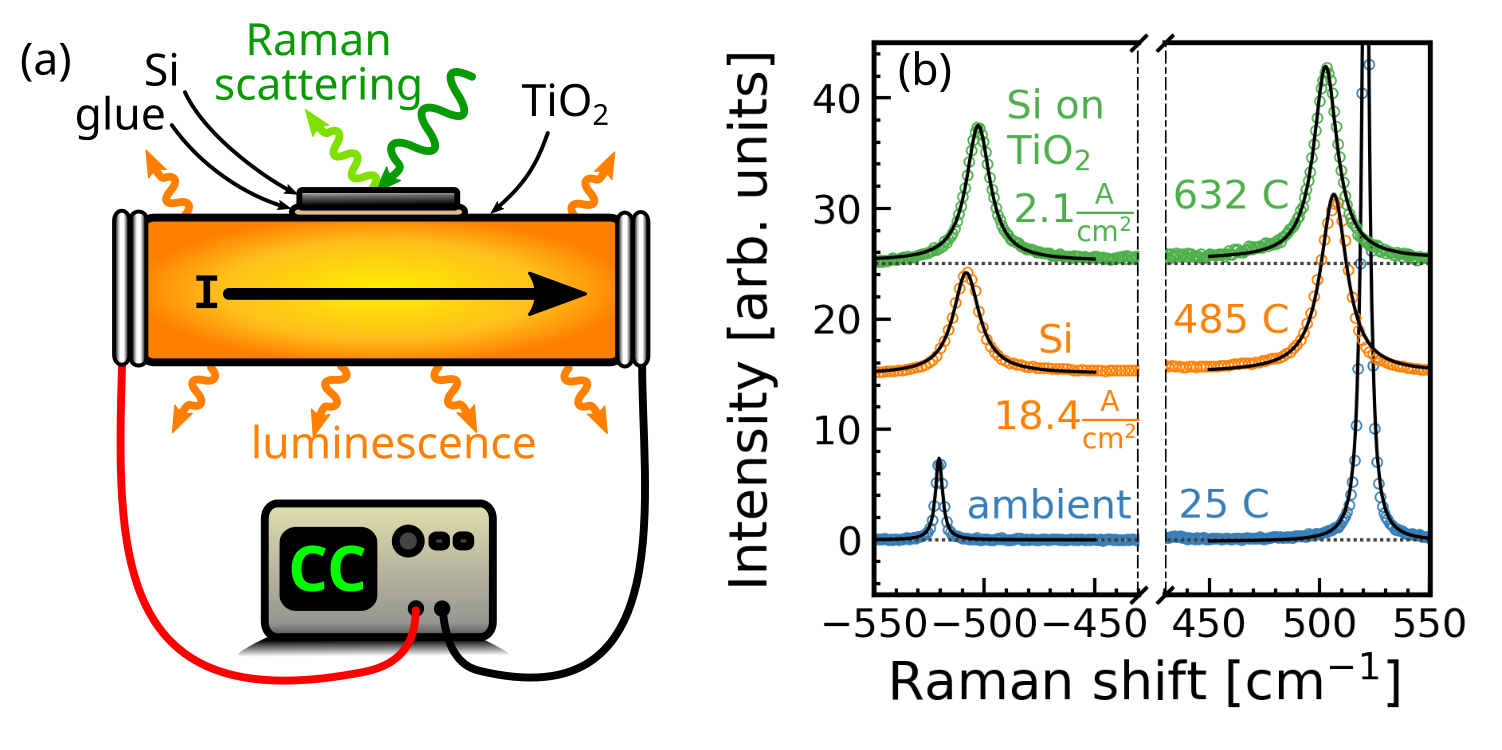}
    \caption{\textbf{Stokes/anti-Stokes Raman thermometry.}  (a) Diagram of the in-situ Raman measurements for TiO$_2$: a small silicon flake was glued to TiO$_2$ and the temperature was measured on the flake. (b) Background-subtracted Raman spectra of silicon phonon offset for clarity: In-situ anti-Stokes (left) and Stokes (right) spectra of flashing silicon (orange) and the same from silicon glued to TiO$_2$ under flash (green). Si spectra with no current at 25 C are in blue. Eq. \ref{eq:detailed-balance} was used to deduce the temperature; we fit the 25 C data to determine "$g$". The fits are represented by solid black lines.}
    \label{fig:raman}
\end{figure}

To perform this analysis we take advantage of the well-characterized intense Raman-active 520 cm$^{-1}$ phonon of silicon (Fig. \ref{fig:raman}(b)). We assume that this phonon is thermalized so that $T_{ph}$ is the same as the sample temperature. When investigating silicon we measured this phonon directly on the sample while current was flowing. For TiO$_2$, we attached a thin $1\times 1$ mm flake of silicon (probe) using electrically insulating high temperature glue and measured the silicon phonon by Raman while passing current through TiO$_2$ (Fig. \ref{fig:raman}(a)). We do this (i) to ensure that detailed balance is valid (which is true in the unpumped silicon probe) and (ii) because the Raman spectrum of TiO$_2$ is too complicated to be useful for extracting temperatures. Excellent agreement of the in-situ phonon energies and widths with the literature \cite{hart1970temperature} confirmed the temperatures we extracted from detailed balance. 

We attached leads in a standard four-point configuration to monitor resistivity of in-situ silicon during the Raman temperature measurements. For TiO$_2$, using a four-point configuration was too technically challenging in the presence of the silicon flake. Rather, we measured the in-situ temperature as function of current density in a two-point setup and then, separately, performed four-point measurements to measure the resistivity as a function of current density (Fig. \ref{fig:si_flash}(c)). We fit the temperature of in-situ TiO$_2$ (dashed line in Fig. \ref{fig:si_flash}(d)) and used this relationship to establish the resistivity vs. temperature, $\rho(T)$, relationship in Figure \ref{fig:tio2_flash}(b). The equilibrium resistivities of pristine silicon and TiO$_2$ were measured in a four-point configuration in a furnace. 

\subsection{Bistability calculations}

The electrostructural bistable behavior of TiO$_2$ is described by the steady-state solutions of the balance equations, Eq. \ref{eq:energy_balance_oxide} under constant voltage (CV) and Eq. \ref{eq:energy_balance_oxide_CC} under constant current (CC). The parameter space for these equations is large: ($V$, $G_0$, $K_0$, $T_b$, $E_0$, $\zeta$). $V$ and $T_b$ are "knobs" that we can turn in the lab; $\zeta$, $G_0$, and $K_0$, are properties of the material. $T$ and $n$ are determined by the response of the material; we look for their solutions at fixed sets of the parameters. It is beneficial to solve "non-dimensionalized" versions of these equations and deduce only the needed parameters by comparison to experiment. 

We define the defect formation temperature scale $T_0 = E_0/k$ and the non-dimensional sample and bath temperatures $x=T/T_0$ and $y=T_b/T_0$ respectively. The CV steady-state energy balance equation becomes $0 = V^2(G_0/K_0T_0^4 ) e^{-1/x + \zeta V / E_0 x} + y^4 - x^4$. We introduce the voltage scale $V_0 = \sqrt{K_0T_0^4/G_0 }$ and non-dimensional voltage $v = V/V_0$ so that the non-dimensional CV steady-state energy balance equation is 
\begin{equation}
\begin{gathered}
    0 = v^2 n + y^4 - x^4 \\
    n = e^{-1/x+vz/x}
    \label{eq:nondim_CV_equation}
\end{gathered}
\end{equation}
with $z = \zeta/\zeta_0$ and $\zeta_0 = E_0 /V_0$ the effective charge scale. 

Ohm's law is $V = I R$. In our model, $v = I R_0 / V_0 n$. We define the current scale $I_0 = V_0 / R_0$ and non-dimensional current $j=I/I_0$ so that $v = j /n \Rightarrow j = vn$. The dimensionless conductance is $g = n$; the dimensionless resistance is $\rho = 1/g = 1/n$. Then the CC steady-state energy balance equations are
\begin{equation}
\begin{gathered}
    0 = \frac{j^2}{n}  + y^4 - x^4 \\
    n = e^{-1/x + jz / nx} .
    \label{eq:nondim_CC_equations}
\end{gathered}
\end{equation}
The set of parameters is $V_0$, $I_0$, and $T_0$ and the rest can be deduced from these. Eq. \ref{eq:nondim_CV_equation} can be solved for $x$ by direct numerical root finding. $n$ is then straightforward to calculate. Solving Eq. \ref{eq:nondim_CC_equations} is slightly more subtle since $n = e^{-1/x + jz / nx} $ is a fixed point equation. We solve eq. \ref{eq:nondim_CC_equations} by filling a dense array of points $0\leq n \leq 1$. We then solve the non-dimensional CC energy balance equation for each $n$ and $j$ for $x$. This determines $n(x,j)$. The solutions of the CV equations, Eq. \ref{eq:nondim_CV_equation}, and CC equations, Eq. \ref{eq:nondim_CC_equations}, are essentially the same as determined by Ohm's law with a key difference: solutions of the CV equations for $j(v)$ can be multivalued while the inverse, $v(j)$, are always single valued. 

We estimate numerical values for the parameters $V_0$, $I_0$, and $T_0$ by comparing our model calculations to experimental results in Figures \ref{fig:si_flash}(d) and \ref{fig:tio2_flash}(a-c). For TiO$_2$, we choose intensive quantities $V_0=400$ V/cm, $I_0=2500$ A/cm$^2$, and $T_0=5960$ K.  We then deduce $E_0\approx 513$ meV, $R_0 = V_0/I_0 = 0.16$ $\Omega$-cm, and $\zeta_0\approx1.3\times10^{-3}$ e (fundamental charge). While our model is crude and we don't expect good quantitative agreement, this $E_0$ is a reasonable estimate for the formation energy of defects in oxides \cite{morgan2010intrinsic,lee2012calculation}. For $z=0.5$ (cf. Figs. \ref{fig:tio2_flash}(c-d)) and for $V=500$ V, which is a typical value in a flash experiment, $V\zeta\approx320$ meV which is also reasonable \cite{abdelouahed2015relevance}. 

For silicon, we fix $\zeta=0$ e and interpret $E_0$ as the activation energy. We estimate $V_0=50$ V/cm, $I_0=1000$ A/cm$^2$, and $T_0=2300$ K by comparing to the current/field in Fig. \ref{fig:calc_results}(a-b) and to the temperature dependent resistivity in Fig. \ref{fig:si_flash}(b). Then $R_0=0.05$ $\Omega$-cm and, with $y=0.1$, $T_b\approx -43^\circ$C. Compared to absolute zero, $-273^\circ C$, this number is reasonable. Just like with TiO$_2$, we don't expect good quantitative agreement. However, we expect to observe runaway at lower temperature in silicon than TiO$_2$ and this is consistent with that. Moreover, we find $E_a\approx198$ meV which is reasonably close to the expected value for silicon, $\approx 500$ meV. 

We estimate the experimental defect concentration from the measured resistivity in TiO$_2$. In our model, resistivity is $\rho=R_0/n$ with $n$ the dimensionless defect concentration. $R_0$ is estimated above. The lattice constants of rutile TiO$_2$ are $a=4.6 ~ \textrm{\AA}$, $b=4.6 ~ \textrm{\AA}$, and $c=3.0 ~ \textrm{\AA}$. The unitcell volume is $\mathcal V = 63.5 ~\textrm{\AA}^3$. We assume a single electron is ionized from each defect and that there is single defect site per unitcell. Then there are $1.6\times10^{22}$ defects sites per cm$^3$ (i.e. $1.6\times10^{22}$ possible carriers/cm$^3$). When $n=1$, all sites are occupied and there are $1.6\times10^{22}$ defects/cm$^3$; when $n=0.1$, there are $1.6\times10^{21}$ defects/cm$^3$ and so on.

\paragraph{Acknowledgments}
We thank Hope Whitelock and Minhyea Lee for technical support and for loaning us equipment for the transport measurements. 

\paragraph{Funding Statement}
All experimental work and related analysis, with the exception of Raman scattering was supported by the U.S. Department of Energy, Office of Basic Energy Sciences, Office of Science, under Contract No. DE-SC0024117. Raman scattering measurements and analysis were supported by the NSF grant No DMR-2210126. RN was supported by the U.S. National Science Foundation under NSF Grant No DMR-2516302. 

\paragraph{Author Contributions}
T.C.S. performed the measurements, theoretical work, and calculations and co-wrote the manuscript. D.A. and T.E.G. assisted with measurements. R.N. advised the theoretical work. D.R. conceived and advised the project and co-wrote the manuscript. 

\paragraph{Conflicts of Interest Disclosure} 
The authors report no conflicts of interest. 

\paragraph{Data Availability}
All data are available from the authors on reasonable request.

\bibliographystyle{elsarticle-num}

\end{document}